\documentstyle[12pt,epsf]{article}
\title{Kinetic Arrest Originating in Competition between\linebreak Attractive Interaction
and Packing  Force}
\author{G.Foffi$^+$, E.Zaccarelli$^+$, F.Sciortino$^*$, P.Tartaglia$^*$,
 K.A.Dawson$^+$\\
\small {\it $^+$ Irish Centre for Colloid Science and Biomaterials, Department of
Chemistry, }\\ 
\small{\it University College Dublin, Belfield, Dublin 4, Ireland }\\ \\
\small{\it $^*$ Dipartimento di Fisica, Universit\`{a} di Roma La Sapienza}\\
\small{\it and  Istituto Nazionale di Fisica della Materia, Unit\`{a} di Roma La Sapienza}\\
\small{\it Piazzale Aldo Moro 2, 00185 Roma, Italy}
\\
({\it Journal of Statistical Physics (in press), 1999})}

\date{}
\begin{document}
\newcommand{\be}{\begin{equation}}
\newcommand{\ee}{\end{equation}}
\newcommand{\barr}{\begin{array}}
\newcommand{\earr}{\end{array}}
\newcommand{\bea}{\begin{eqnarray}}
\newcommand{\eea}{\end{eqnarray}}

\maketitle
\begin{abstract}
We discuss the situation where attractive and repulsive portions of the
inter-particle potential both contribute significantly to glass formation. We
introduce the square-well potential as prototypical model for this situation, and
{\it reject} the Baxter as a useful model for comparison to experiment on glasses, based on our treatment within mode coupling theory. We present explicit
result for various well-widths, and show that, for narrow wells, there is a useful
analytical formula that would be suitable for experimentalist working in the field of
colloidal science. We raise the question as to whether, in a more exact treatment,
the sticky sphere limit might have an infinite glass transition temperature, or a high but finite one.   
\end{abstract}
\section{Introduction}
Kinetic arrest phenomena at a transition temperature $T_{c}$  occurring just
prior to the thermodynamic glass transition, $T_{g}$, are well known and have
been studied extensively using the mode-coupling theory and simulation methods.
Without being exhaustive, we can point to a number of reviews
\cite{gotze,kawasaki} and papers \cite{gotze2,gotze3,kob} in the literature. 
There has also been considerable experimental work on this field
\cite{vanmegen1}-\cite{halalay}.
Typically such phenomena are considered to be driven primarily
by packing effects where the repulsive  part of the potential is of primary
importance, with the attraction providing merely a modulation of the overall
phenomena. In this case the hard sphere system exhibits most of the relevant
phenomena and may be viewed as the prototypical model of this type of kinetic
arrest \cite{barrat}.\\
However, there has recently been interest \cite{sciotar,bergen} in cases
where the attractive part of the potential, or rather its interplay with repulsion, is
more deeply implicated in the arrest phenomenon. This can lead to glasses with much
richer structure, including long-ranged density correlations that are frozen
into the system. 
New dynamical phenomena might also be expected in such systems,
especially where there is subtle interplay between attraction and repulsion near
the glass transition.
As in the repulsive case, it is natural to seek a prototypical model that can be
conveniently studied within mode-coupling theory with the aim of elucidating the
special features of such a system. It is in particular desirable to consider
those aspects that represent more than simple modulations of hard-core
behaviour. One such choice is represented by the Baxter model \cite{baxter1} which
 has been
studied by some of the authors \cite{sciotar,bergen} in  previous publications
using the Percus-Yevick approximation \cite{percus} for the static structure factor,
 and the
mode-coupling theory. The results are promising in that novel dynamical
phenomena do emerge, and also it is possible to create kinetic arrest where a
long correlation length scale is quenched into the system.\\
Of course, some of the limitations of a model with hard-core potential and
delta-function like attractive potential, as treated by Percus-Yevick \cite{percus} are
evident. For example, it is believed that such a potential would lead only to
disordered and crystalline phases, rather than the liquid-gas phase separation
implied by the P-Y approximation \cite{stell1}. 
On the other hand, it may still be possible to
make some progress with this approach. For example, in the disordered phase the 
P-Y approximation is reasonably good for the structure factor \cite{caccamo}, so in
those cases where the kinetic arrest arises, prior in temperature or time to a
crystallization, we may expect reasonable results.
It may even be the case that the capability to study arrest near what is
believed to be a `metastable' liquid-gas transition may be of some practical
interest \cite{lekke1, lekke2, lekke3, verduin}. Here we discuss some 
unsatisfactory features originating from the large-$q$-tail of the static 
structure factor, in conjunction with the study of the ideal glass transition 
in the frame of the mode coupling theory (MCT). \\ 
It is perhaps prudent to mention at this stage that the use of Mode
Coupling Theory (MCT) for such questions is itself not without controversial
aspects. There are numerous criticisms of MCT, and in any case, even if one does
accept it as a viable mean-field-type theory, there remains the criticism that
it neglects such effects as `hopping' and other phenomena \cite{kawasaki2}. 
These may be relevant in square-well systems.\\
However, we shall further argue that, within the MCT, the P-Y approximation and 
possibly even the exact structure factor of the Baxter model, leads to 
very high or infinite glass transition temperature for a broad range of 
packing fractions in the region where  attractive interactions dominate. 
The outcome of our deliberations shall be that
we shall propose as our prototypical model  the `narrow' square well potential,
treated by P-Y and MCT. We shall show that, for this case,  the structure factor 
is damped at large-$q$ due to the finite well width, and that this leads to
finite values of $\tau_c$, the effective transition temperature, throughout the phase-diagram. Moreover, the 'narrow' square well potential seems to share most of the feature of the kinetic arrest discussed in earlier works \cite{sciotar, bergen}. 
We shall conclude that the underlying phenomena reported there are therefore 
\emph{robust} and likely to be observed experimentally, even though the Baxter
model is flawed.\\
The organization of these observations is as follows.
We shall first consider 
the static structure factor of the square well potential as the input to the
MCT. This will permit us to explain new features introduced by a finite
well-width.
Next we shall consider the effects of these features on the MCT prediction of 
the transition.
Finally we shall discuss the general behaviour of the arrest phenomenon as a
function of well-width.

\section{The P-Y Approximation to the Square-Well Potential}
\begin{figure}[h]
\centerline{
            \epsfxsize=9.0cm
				\epsfbox{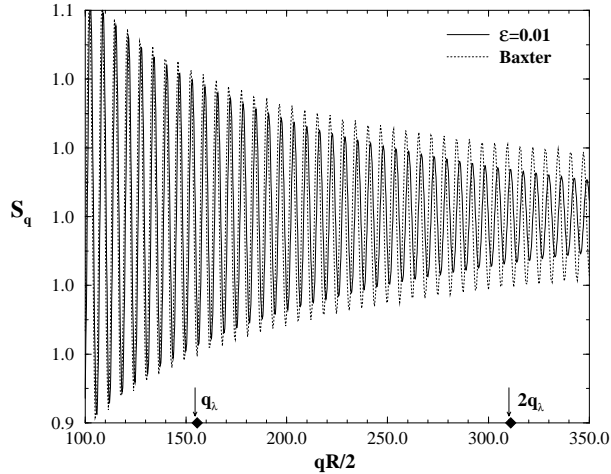}
		     }
			  
				
\caption{Structure factor at large-$q$ of the Baxter model and  
$\epsilon$-expansion for $\epsilon=0.01$. The
value $q_{\lambda}\!\!=\!\!\frac{\pi}{\epsilon R'}$ and by the examination of
 Fig.\ref{taug} it
is evident that much beyond this point there is little contribution to the
transition temperature.}
\label{eps0.01&Baxter}
\end{figure}

We consider a square well potential with repulsive core of diameter $R$ and total
range $R'$. The well width is therefore parametrized by $\epsilon=
\frac{R'-R}{R'}$, and the well depth by $u$. Note that by taking the  
appropriate limit of well
depth to width we may recover the Baxter model which possesses an analytical
solution for the structure factor.
By the Baxter sticky sphere model \cite{baxter1} we mean $\epsilon \rightarrow0$ and the well
depth becomes infinite according to $\frac{1}{\beta}\ln(12\tau\epsilon)$, $\tau$
being an effective temperature and $\beta=\frac{1}{k_B T}$ .
To solve the Baxter model within PY an external
parameter $\lambda$ is introduced to define $h(r)$ within the range of the 
attraction. It is then easily calculated through a second order
equation arising from the matching conditions that have to be verified. 
Since the limit $\lambda \rightarrow 0$ corresponds to the hard sphere limit, 
it may be thought as an attractive energy scale.\\
Although no analytic  solution exists for the
square well problem, this model potential  is  simple enough to be exhaustively
studied.
Thus the integral equation for the square well problem is readily derived using  Baxter's approach. The equations are
\be
\label{numeric1}
r c(r) = -Q'(r) + 2 \pi \rho \int_{r}^{R'} dt Q'(t)Q(t-r) 
\ee
for $0 < r < R'$, and
\be r h(r) = -Q'(r) + 2 \pi \rho \int_{0}^{R'} dt (r-t) (h(|r-t|) Q(t)
\label{numeric2}
\ee
for $r > 0$, where $c(r)$ is the direct correlation function, 
$h(r)$ is the indirect correlation function and  
$Q'$ is the derivative of $Q(r)$ \cite{bax}.
The function $Q(r)$ is related to $S(q)$ via
\be
S(q)^{-1}=\tilde Q(q) \tilde Q(-q)
\label{numeric3}
\ee
and
\be
\tilde Q(q) = 1 - 2 \pi \rho \int_{0}^{R'} dr e^{i q r} Q(r).
\label{numeric4}
\ee
This equation can be solved numerically using the P-Y approximation as 
closure, and it is via this expression that the potential enters the
formulation.
Of course, the PY closure has many limitations. However, it is quite
acceptable for hard spheres, and it is expected to be reasonable for
short-ranged potentials \cite{caccamo}.\\
Instead of a numerical solution, the P-Y theory  may be solved in a series expansion for small
well-width, $\epsilon$,a program that was partially carried out by Menon et al.
\cite{manon}.
The result for the structure factor is,
\bea
\label{chen} 
\frac{1}{S(q)}-1&=&24\eta \left[ \alpha f_{2}(\bar{q})+\beta
f_{3}(\bar{q})+\frac{1}{2}\eta\alpha
f_{5}(\bar{q})\right]+4\eta^{2}\lambda^{2}\epsilon^{2}\left[f_{2}
(\epsilon\bar{q})-\frac{1}{2}f_{3}(\epsilon\bar{q})\right]+ \nonumber \\
&+&2\eta^{2}\lambda^{2} 
\left[f_{1}(\bar{q})-\epsilon^{2}f_{1}(\epsilon\bar{q})\right]-\frac{2\eta\lambda}
{\epsilon}\left[f_{1}(\bar{q})-(1-\epsilon)^{2}f_{1}((1-\epsilon)\bar{q})\right]
- \nonumber \\ 
&-&24\eta\left[f_{2}(\bar{q})-(1-\epsilon)^{3}f_{2}((1-\epsilon)\bar{q})\right]
\eea
where $\bar{q}=qR'$, $\eta=\frac{\pi}{6} \,\rho R'^{3}$ and 
$\alpha$, $\beta$ and $\mu$ are given by
\bea \alpha&=&\frac{(1-2\eta-\mu)^2}{(1-\eta)^4}\nonumber \\
     \beta&=&-\frac{3\eta(2+\eta)^{2}-2\mu(1+7\eta+\eta^{2})+\mu^{2}(1+\eta)}{2(1-\eta)^4} \\
	  \mu&=&\lambda \eta(1-\eta) \nonumber
 \label{alfabeta}
\eea
and  $f_{n}(x)$ are defined as:
\bea f_{1}(x)&=&\frac{1-\cos{x}}{x^{2}}, \nonumber \\
     f_{2}(x)&=&\frac{\sin{x}-x\cos{x}}{x^{3}}, \nonumber \\
     f_{3}(x)&=&\frac{2x\sin{x}-(x^{2}-2)\cos{x}-2}{x^{4}}, \nonumber \\
     f_{5}(x)&=&\frac{(4x^{3}-24x)\sin{x}-(x^{4}-12x^{2}+24)\cos{x}+24}{x^{6}}.
\eea
a formula that was derived by Liu et al. \cite{chen}. Note that $\eta$ is
the effective volume fraction calculated according to the large length, $R'$, that defines
the full range of the potential. Sometimes we shall also use the real volume
fraction calculated with $R$, the hard core diameter, and this is named $\phi$.
The two quantities are the same in the limit of Baxter spheres.\\
In fact, one can push this asymptotic theory a little further than has been done previously. That is, we can  solve all parts of the theory to order $\epsilon$, including the $\lambda$-equation, obtaining the following,
\bea \lambda^2\left(1+\frac{3\epsilon\eta}{1-\eta}\right)
&-&12\lambda\left[\frac{\tau}{\eta}+\frac{1}{1-\eta}-
\epsilon\frac{1-\frac{11}{4}\eta+\eta^2}{(1-\eta)^2}\right]+ \nonumber \\ 
&+&\frac{6}{(1-\eta)^2}\left(\frac{2}{\eta}+1-12\epsilon(1-\eta)\right)=0
\label{eps,eq,lambda}\eea
where the physical solution $\lambda(\epsilon)$ is now of the form:
\begin{eqnarray} \lambda(\epsilon)&=&6\left[\left(\frac{1}{1-\eta}+\frac{\tau}
{\eta}\right)\left(1-\frac{3\epsilon\eta}{1-\eta}\right)-\epsilon
\frac{1-\frac{11\eta}{4}+\eta^2}{(1-\eta)^2}\right]  \nonumber \\ 
&\pm &\left\{ 36\left[\left(\frac{1}{1-\eta}+\frac{\tau}
{\eta}\right)\left(1-\frac{3\epsilon\eta}{1-\eta}\right)-\epsilon
\frac{1-\frac{11\eta}{4}+\eta^2}{(1-\eta)^2}\right]^{2}-\right.\nonumber\\ 
&-&\left.\frac{6}{(1-\eta)^2}\left[\left(\frac{2}{\eta}+ 
1\right)\left(1-\frac{3\epsilon\eta}{1-\eta}\right)-12\epsilon(1-\eta)\right]
\right\}^{1/2} \label{eps,sol,lambda} \end{eqnarray}
It is fairly straightforward then to compare the results of equation (\ref{chen}) where $\lambda=\lambda(\epsilon)$ with numerical solution 
of equations(\ref{numeric1}, \ref{numeric2}, \ref{numeric3}, \ref{numeric4}). 
It is also possible to illustrate differences between the Baxter 
sticky
sphere and the square well model.
We shall here present a representative selection of results from three
values of $\epsilon$ ($\epsilon=0.01$,$\epsilon=0.03$ and $\epsilon=0.09$), 
chosen to illustrate various aspects of the mode-coupling calculations.
The smallest value of $\epsilon$ is representative of {\em sticky-sphere} type
behaviour, but with a finite temperature transition. The largest value
($\epsilon$=0.09) already exhibits properties closer to that of a square well.
\begin{figure}[h]
\centerline{
            \epsfxsize=10.0cm
				\epsfbox{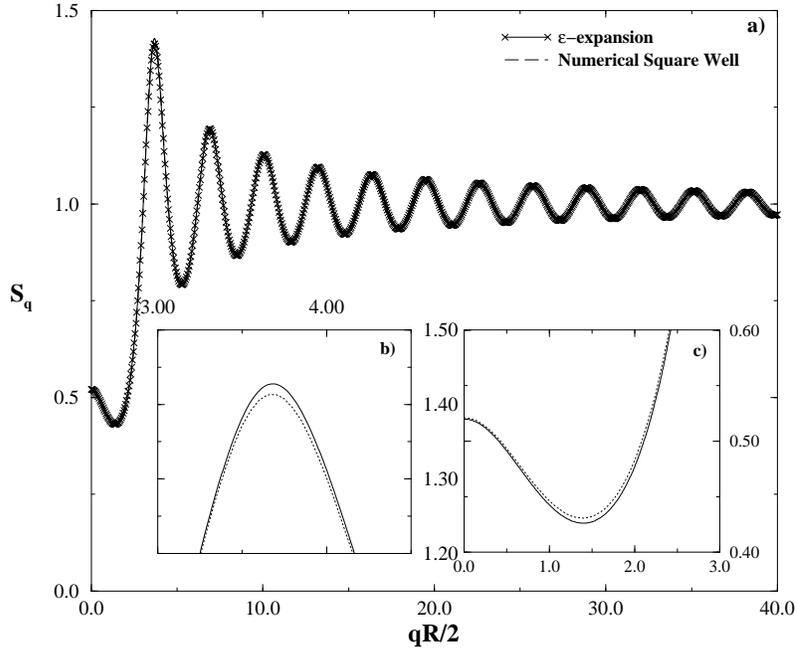}            
				}
\caption{ a) Square well numerical result plotted versus $\epsilon$-expansion for$\epsilon=0.01$.
  b) Detail of the first peak of the structure factor (the
dotted line is the numerical calculated  square well, the continuous line is the $\epsilon$-expansion).
c) Detail of the {\em small}-q region of the structure factor.
        }
\label{eps0.01&SW}
\end{figure}
We note first that, for $\epsilon=0.01$, the agreement between the numerical
calculation of equations (\ref{numeric1}, \ref{numeric2}, \ref{numeric3}, \ref{numeric4}) and the leading order in the well-width
expansion is essentially perfect (Fig.\ref{eps0.01&SW}), and we may accept 
equation (\ref{chen}) as defining the  P-Y approximation to the small well-width
problem. As one might expect for such small $\epsilon$, the moderately small-$q$ (corresponding to the nearest neighbour distances) behaviour of
the structure factor in equation (\ref{chen}) is  in good agreement also with the
Baxter model. However there are some very important differences
 for the large-$q$
behaviour of  $S(q)$ as illustrated in Fig. \ref{eps0.01&Baxter}. There are two
points to note here that relate to later MCT calculations. Firstly the MCT
calculations are sufficiently delicate at large-$q$ for it to be necessary to
include many momentum points on the numerical grid. It is therefore highly
advantageous to have a formula such as equation (\ref{chen}), rather than a
numerical solution. 
Secondly from Fig.\ref{eps0.01&Baxter} (and later comparison of Fig.\ref{taug}) we
 see that beyond $q_{\lambda}$, the structure factor of the square well model 
 decreases more rapidly than that of the Baxter model. 
In fact consideration of finite well-width leads, for many
purposes, to the effective introduction of a {\em cut-off} at
twice the well width,
\be
\label{cutoff}
q_{\lambda}=\frac{\pi}{(R'-R)}.
\ee
\begin{figure}[h]
\centerline{
				\epsfxsize=9.0cm
				\epsfbox{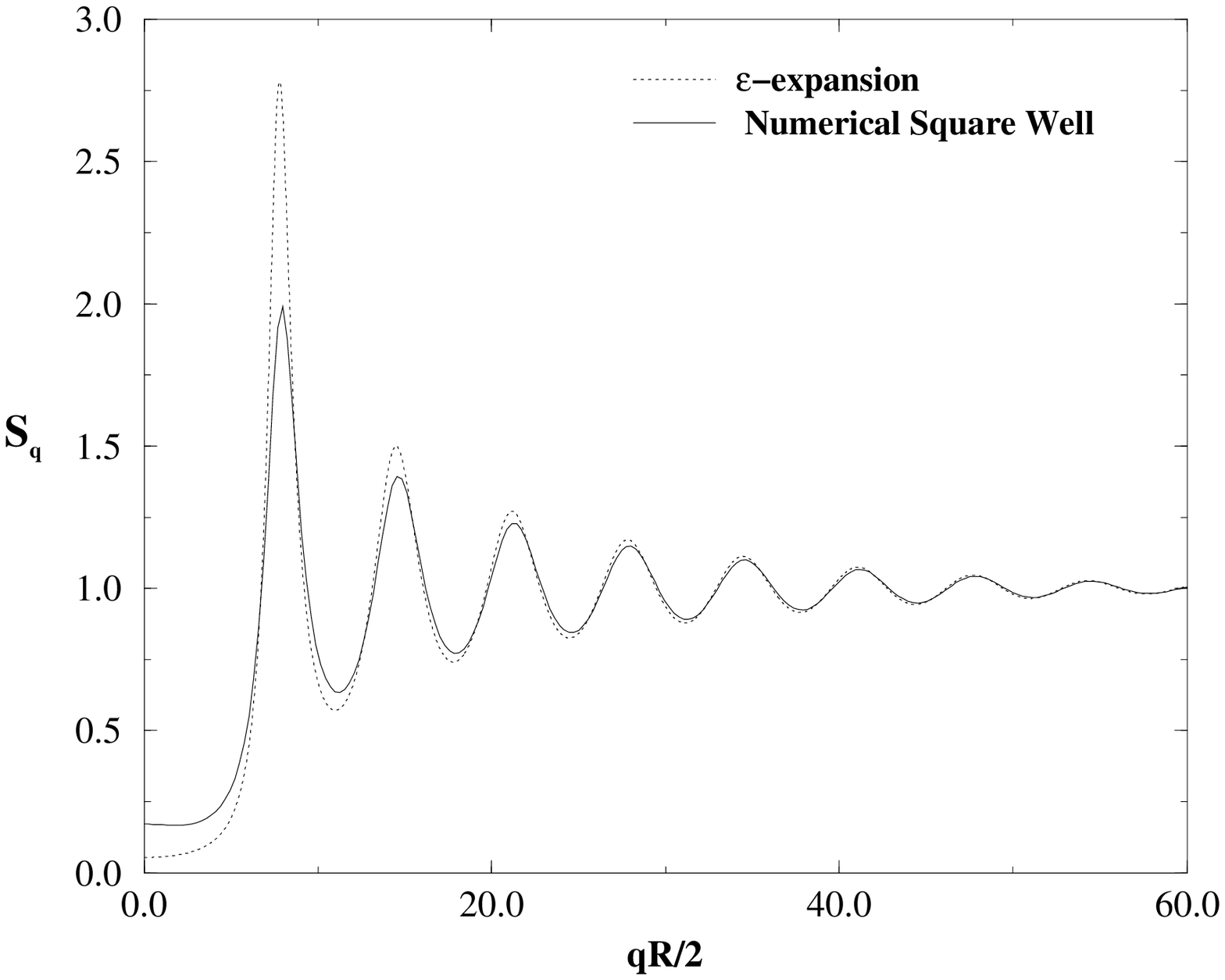}
				}
				
\caption{Square well versus $\epsilon$-expansion for $\epsilon=0.09$.}
\label{ciao}
\end{figure}

It is this reduction of the range of $S(q)$ that leads ultimately to meaningful
mode-coupling calculation.\\
Further comparisons between the $\epsilon$-expansion and the numerical results from
equation (\ref{numeric1}) are also possible.  As might be expected, at 
$\epsilon=0.09$ the deviation between the two results becomes
significant (Fig.\ref{ciao}), both for
small-$q$ and the first peak, and thereafter one must rely on the numerical
solution. Even so, it is interesting to note that the large-$q$ behaviour is still quite well
represented by equation (\ref{chen}).\\
We now turn to the thermodynamic aspects of the P-Y solution of the finite well
problem. We have noted earlier some intrinsic failings of the P-Y approximation of
the Baxter model in that liquid-gas rather than the expected crystallization 
transition is produced \cite{stell1}.
There are other unusual features of
the P-Y solution, such as the asymmetry of the phase diagram with respect to the
spinodal (see for example \cite{fisher,gallerani1,gallerani2}) 
 and the uncertain status of the phase diagram to the left of the critical point.
In fact, a careful analysis of the expansion in square-well width leads to a few
new features, not present in the Baxter solution. Amongst them there is the observation
that the phase diagram is no longer so asymmetric with respect to the critical
point, and there is a finite region of real solution beneath the spinodal curve,
near the critical point\cite{zacca}. These details need not concern us here, since
we shall be interested in points of the phase diagram above the P-Y phase
separation.\\For our present purpose it is possible to use
the P-Y structure factors for temperatures and densities to the stable (right) side
of the high density branch of the spinodal line. \\
In summation, then, we believe that the P-Y approximation, and in particularly the
leading order of the {\em small well-width} expansion given by equation (\ref{chen})
is, at least, a useful approximation to the structure factor for an interesting range of
parameters of the square well potential. We shall now study thus range within MCT.

\section{The Mode Coupling Theory}

\begin{figure}
\centerline{
				\epsfxsize=9.0cm
				\epsfbox{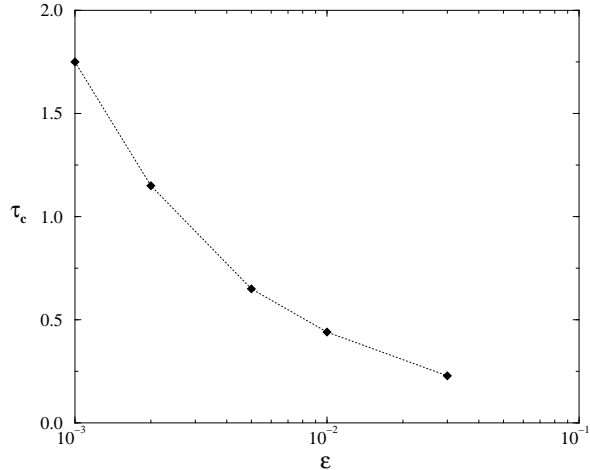}	}
\caption{Behaviour of $\tau_{c}$ in function of $\epsilon$ at $\phi$=0.2}
\label{epsi}
\end{figure}

The mode coupling theory provides a description of the kinetic arrest phenomenon. For certain
values of the interaction parameters the density-density correlation function
\be
\phi_{q}(t)=\frac{< \!\delta \rho^{*}({\bf q},t)\delta \rho({\bf q},0)\!>}
{NS_{q}}
\label{correlation}
\ee
possesses a long-time decay with a non-zero infinite time limit. When this
occurs,  diffusion slow dramatically and the viscosity diverges: the glass
transition occurs. The Zwanzig and Mori \cite{mori} formalism and MCT ideas
 \cite{gotze} lead to the equations
\be
\label{GLE}
\ddot{\phi_{q}}(t)+\Omega_{q}^{2}\phi_{q}(t)+\nu_{q}\dot{\phi_{q}}(t)+
\Omega_{q}^{2}\int_{0}^{t}m_{q}(t-t')\dot{\phi_{q}}(t')dt'=0
\ee

where  $m_{q}(t)$ is given by
\be
\label{memory}
m_{q}=\frac{1}{2}\int \frac{d^{3}k}{(2\pi)^{3}}V( {\bf q},{\bf k})\phi_{k}(t)
\phi_{|{\bf q}-{\bf k}|}(t)
\ee
the  vertex function is
\be
\label{vertex}
V( {\bf q},{\bf k}) = \frac{\rho}{ q^{4}}\left({\bf q}
\cdot( {\bf q} - {\bf k}) 
c_{|{\bf q}-{\bf k}|}+ {\bf q}\cdot {\bf k} c_{k}\right)
^{2}
S_{q}S_{k}S_{|{\bf q}-{\bf k}|}
\ee
and the two quantity $\Omega_{q}$ and $\nu_{q}$ are respectively the 
 characteristic frequency and a  white noise term due to the  fast
 part of the memory function; they are defined as,
\bea
\Omega_{q}&=&\frac{q^{2}k_{B}T}{mS(q)} \nonumber \\
\nu_{q}&=&\nu_{1}q^{2} \nonumber 
\eea
and $\nu_{1}=1$ in our calculations.
In the limit $t \rightarrow \infty$ we find the equation for the  static problem
\be
\frac{f_{q}}{1- f_{q}} = \frac{1}{2} \int \! \! \frac{d^{3}k}{(2\pi)^{3}} \,\,
 V( {\bf q},{\bf k})f_{k}f_{|{\bf q}-{\bf k}|}  
\label{mct}
\ee

where $f_{q}$ is the {\em Edwards-Anderson} factor:
\be
\label{nonergo}
f_{q}= \lim_{t\rightarrow \infty} \phi_{q}(t),
\ee
$S_{q}$ is the static structure factor and $c_{q}=\frac{S_{q} -1}{\rho S_{q}}$ is
 the
Fourier transform of the direct correlation function following the Ornstein-Zernike
relation.\\
It is clear that $f_{q} = 0$ is always a solution of the equation
(\ref{nonergo}). In fact it is possible to show \cite{gotze} that, where the potential
is purely repulsive and at low densities, 
this is the stable solution, and the system is in the liquid phase. Above a certain
$\phi_{c}$ (that depends on the parameters of the system ) non-zero solutions begin
to appear and  non-ergodic behaviour appears. 
The   ideal glass line, moreover, can be defined studying the
behaviour of the eigenvalues of the  stability matrix of the system
\cite{gotze}.
Here we have bracketed the transition line by iteratively solving equation
(\ref{mct}). Where more precise results have been required  we have studied the stability
matrix. Of course, all these calculations are subject to numerical error,
particularly with respect to the numerical momentum cut-off in the integrals.
Since the attractive glass requires consideration of both repulsive and attractive
parts of the potential, the former length scale being set to be the well-width, we
have carried out finite-size analysis of our results to check their validity.
Now we are in a position to solve the kinetic arrest temperature, $\tau_{c}$, as a
function of well parameters and the structure factor. From now on we shall
refer to the {\em effective} temperature $\tau$ as introduced by Baxter and
related (for small $\epsilon$) to the temperature via: $\tau=\frac{1}{12\epsilon}exp{\frac{ u}{k_{B}T}}$
\cite{manon}.  
The first point to note is that, for the Baxter model in the region where
attractions are relevant (previously labelled as $B_{2}$ in  references
 \cite{sciotar} and \cite{bergen}), there are, strictly speaking, very 
 substantial contributions from the 
long tail of the Baxter solution, not taken into account in earlier works.
 The apparent transition observed previously  is a consequence
of truncating the long tail of the Baxter solution. Indeed, we find that
$\tau_{c}$ rises  with the number of $q$-vectors taken into account
in the calculation.\\
\begin{figure}[tbp]
\centerline{
				\epsfxsize=9.0cm
				\epsfbox{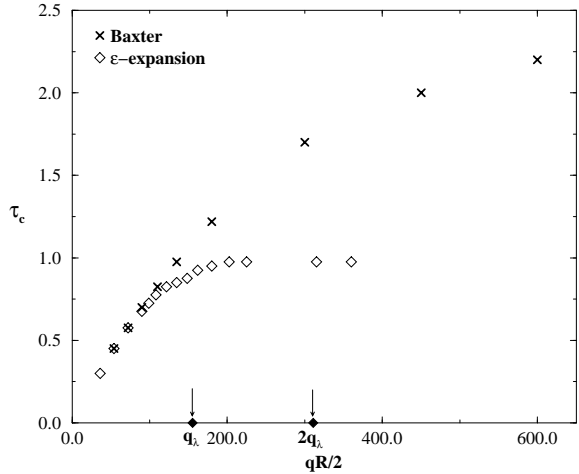}
				}
\caption{Behaviour of $\tau_{c}$ in function of the cut-off as evaluated from
equation (\ref{mct}). $\epsilon$=0.01 and $\phi$=0.3881.
As before $q_{\lambda}=\frac{\pi}{\epsilon R'}$. }
\label{taug}
\end{figure}
For example, in the case of Fig.\ref{taug} ($\epsilon$=0.01 and $\phi$=0.3881) we see
that the $\tau_{c}$ continues to change with the number of points in the numerical
grid  until we
reach a well-width cut-off equal to twice the well width ($\frac{\pi}{\epsilon R'}$).Of
course, the cut-off required is somewhat dependent on the property of the system
chosen. 
However the general outline of the observation is clear. The attractive glass
transition is strongly {\em finite-size} dependent due to the contribution to
the integral equation (\ref{mct}), from momentum scales up to around
$q_{\lambda}\!\!=\!\!\frac{\pi}{\epsilon R'}$. Thus, for very small well widths
the calculations require many thousands of points, and for the Baxter model
there appears to be no useful number of points that can ensure convergence of
the transition temperature in the attractive region of the transition. The
implication is clear. It is the (small) finite-well-width model that is the
appropriate prototype of the system we discuss rather then the literal Baxter Model.
However, we make another observation, illustrated quite well by Fig.\ref{taug}. That
is, paradoxically, the Baxter model solution can be applied in the mode coupling
theory to study attractive glasses providing the appropriate well-width cut-off
($q_{\lambda}$) is applied. We can see this because $\tau_c$ from the Baxter solution
truncated at $q_\lambda$ is comparable to $\tau_c$ from the true P-Y square well
potential of width $\epsilon=\frac{\pi}{q_{\lambda}R'}$. In some cases in the
literature we have therefore the interesting situation that the Baxter structure
factor appears to work due to numerical cut-off, even though the full model has
an unrealistic glass transition temperature. The discussion is not purely academic since
experimentalist fit data to such models. In future we recommend that equation
(\ref{chen}) be applied up to around $\epsilon$=0.03 for the case of attractive
colloidal particles. Beyond this value of $\epsilon$, more care must be taken.\\
We now turn to another aspect of this question. We have calculated the
asymptotic $\tau_{c}$ as a function of $\epsilon$ for $\phi$=0.2 
(Fig.\ref{epsi}). 
Evidently, as $\epsilon$ decreases the glass temperature rises sharply (due to the
large-$q$ tail) so that at
$\epsilon$=0.001 the effective temperature $\tau_{c}$ is very high, roughly
seventeen times higher than the Baxter gas-liquid critical temperature. 
For numerical calculation,  we are unable to conclusively show the asymptotic limit $\tau_{c}$ 
($\epsilon \rightarrow 0$).
However, we believe that the glass appears to be the stable solution for all effective temperatures where the system does not behave as a herd-sphere fluid.
This is quite a practical question. Colloidal particles with vanishing
well-width, but large well-depths would therefore not  undergo the expected 
fluid-solid transition
\cite{stell1}, but would generally found to be a glass, up to temperature  where
attraction are irrelevant  in any case, and the particles behave as hard spheres. 
Since, for finite $\epsilon$,  we now have finite values of $\tau_c$, it is natural to ask
if the important dynamical features of the model described in reference
\cite{sciotar} are recovered. The existence of a logarithmic decay of density-density correlation, associated to the presence of a cusp singularity in the parameter space in the case of ``narrow'' square well potentials is under investigation.

\section{Conclusions}
As we have noted earlier, it is of interest to identify a simple model, and
treatment of that model that exhibits the principal phenomena where the attractive
part of the potential is important in glass formation. We have seen that one good
candidate is the square well potential where, for very narrow wells,  important new
properties are already evident. Thus, glasses with frozen density fluctuations, including those
where the correlation lengths $\xi$ is quite large, appear beneath the attractive glass
curve. In addition we have seen that well-width is quite an important parameter. Truly sticky
spheres in the Baxter sense, or experimental approximations to these, probably are nearly
always found as glasses, perhaps the powdery precipitates found in particle glasses being
examples of this. For more moderate square-well parameters we may expect 
interesting glass behaviour, and the details of the transition and 
attendant dynamics are under examination. We have also noted that, 
from the experimental point of view, the finite-well structure factor formula 
(\ref{chen}) represents a more realistic approach to the dynamics of such 
system, though indication of this were already clear much earlier
\cite{chen}.\\
Finally, we offer the conjecture that the exact model of Baxter spheres would 
be a glass for any effective temperature which preserves the presence of the 
attractive interactions in the potential. This could be in agreement with
Stell's \cite{stell1} clustering ideas for the Baxter model. In addition,
however, we have illustrated how the situation evolves away from the Baxter
limit. Proof, or modification, of the Baxter limit scenario  
may be an interesting challenge to some of the readers of this paper.


\begin{thebibliography}{99}
\bibitem{gotze} W. G\"{o}tze, in {\it Liquids, Freezing and Glass Transition}, 
ed. J.P.Hansen, D. Levesque and J. Zinn-Justin, Les Houches Session LI, 
1989(North-Holland, Amsterdam, 1991).
\bibitem{kawasaki} K. Kawasaki, in {\it Phase Transitions and Critical
Phenomena}, Vol. 5A, C. Domb and M> S. Green, eds (Academic Press, London,
1976).
\bibitem{gotze2} U.Bengtzelius, W.G\"{o}tze, A. Sj\"{o}lander, {\it J.Phys.c},
{\bf17}, 5915 (1984).
\bibitem{gotze3} W. Gotze and L. Sjogren, {\it J. Phys. Cond. Matt.} {\bf 1},
4203 (1989); {\it Phys. Rev. A} {\bf 43}, 5442 (1991); {\it Rep. Prog. Phys.}
{\bf 55}, 241 (1992).
\bibitem{kob} W.Kob, H.C.Anderson, {\it Phys. Rev E} {\bf51}, 4626 (1995) and
references therein.
\bibitem{vanmegen1} W. van Megen, S. M. Underwood and P. N. Pusey, {\it Phys.
Rev. Lett.} {\bf 67}, 1586 (1991);\\ 
W. van Megen and S. M. Underwood{\it Phys. Rev. Lett.} {\bf 70}, 2766 (1993).
\bibitem{mezei1} F. Mezei, W. Knaak and B. Farago, {\it Phys. Rev. Lett.} {\bf
58}, 571(1987);\\ {\it Phys. Scr.} 
{\bf T19}, 571 (1987).
\bibitem{mezei3} E. Kartini, M. F. Collins, B. Collier, F. Mezei and E. C.
Svensson, {\it Phys. Rev. B} {\bf 54}, 6296 (1996).
\bibitem{richter} D. Richter, B. Frick and B. Farago, {\it Phys. Rew. Lett.} 
{\bf 61}, 2465 (1988).
\bibitem{richter2} B. Frick, D. Richter, W. Petry and U. Buchenau, {\it Z.
Phys. B} {\bf 70}, 73 (1988).
\bibitem{fujara} F. Fujara and W. Petry, {\it Europhys. Lett} {\bf 4}, 921
(1987).
\bibitem{arbe} A. Arbe, U. Buchenau, L. Willner, D. Richter, B. Farago and J.
Colmenero, {\it Phys. Rev. Lett.} {\bf 76}, 1872 (1996).
\bibitem{halalay} E. W\"{o}lfle and B. Stoll, {\it Colloid Polym. Sci.} {\bf
258}, 300 (1980); I. C. Halalay, {\it J. Phys. Cond. Matt.} {\bf 8}, 6157
(1996).
\bibitem{barrat} J. L. Barrat, W. Gotze and A. Latz, {\it J. Phys. Cond. Matt.} 
{\bf 1}, 7163 (1989).
\bibitem{sciotar} L.Fabbian, W.G\"{o}tze, F.Sciortino, P.Tartaglia and F.Thiery,
{\it Phys. Rev. E} {\bf 59},  1347 (1999).
\bibitem{bergen} J.Berghenholtz.M. Fuchs, {\it Phys. Rev E} {\bf 59}, 5706, 
(1999).
\bibitem{baxter1} R.J.Baxter, {\it J. Chem. Phys.} {\bf 49}, 2770 (1968).
\bibitem{percus} J.K.Percus and G.J.Yevick, {\it Phys. Rev.}  {\bf 110 }, 1 (1958).
\bibitem{stell1} G.Stell, {\it J. Stat. Phys.} {\bf 63}, 1203 (1991);\\
B. Borstnik, C.G. Jesudason and G.Stell, {\it J. Chem. Phys.} 
{\bf 106}, 9762 (1997); \\ R.P.Sear 1999 {\it J. Phys. Cond. Matt.} in press
(1999).
\bibitem{caccamo} C.Caccamo, {\it Phys. Rep. } {\bf 274}, 1 (1996).
\bibitem{lekke1} H.N.W. Lekkerker, W.C.K. Poon, P.N. Pusey, A.Stroobants and P.B. Warren, {\it Europhys. Lett.} {\bf 20}, 559 (1992).
\bibitem{lekke2} C.F. Tejero, A. Daanoun, H.N.W. Lekkerker and M. Baus, {\it Phys. Rev. Lett.}, {\bf 73}, 752 (1994).
\bibitem{lekke3} N.A.M. Verhaegh, D. Asnaghi, H.N.W. Lekkerker, M. Giglio and 
L. Cipelletti, {\it Physica A} {\bf 242}, 104 (1997).
\bibitem{verduin} H.Verdun and J.K.GDhont, {\it J.Colloid Interface Sci}
{\bf 172}, 425 (1995).
\bibitem{kawasaki2} K. Kawasaki, {\it Physica A} {\bf 243}, 25 (1997).
\bibitem{bax} R.J.Baxter, {\it Phys. Rev.} {\bf 154}, 170 (1967).
\bibitem{manon} S.V.G. Menon, C. Manohar and K. Srinivasa Rao, {\it  
J. Chem. Phys.} {\bf 95}, 9186 (1991).
\bibitem{chen} Y.C. Liu, S.H. Chen and J.S. Huang, {\it Phys. Rev. E} {\bf
54}, 1698 (1996).
\bibitem{sciotar2} F.Sciortino, P.Gallo, P.Tartaglia, S.-H. Chen, {\it Phys. Rev
E} {\bf54}, 6331(1996).
\bibitem{fisher} S.Fishman and M.Fisher, {\it Physica A} {\bf 108}, 1 (1981).
\bibitem{gallerani1} F.Gallerani, G.Lo Vecchio, L.Reatto, {\it Phys. Rev A}
{\bf31}, 511 (1985).
\bibitem{gallerani2} F.Gallerani, G.Lo Vecchio, L.Reatto, {\it Phys. Rev A}
{\bf32}, 2526 (1985).
\bibitem{zacca} E.Zaccarelli, {\it Thesis }, Universita'  La Sapienza',
Rome, 1999 (to be published).
\bibitem{mori} J.P.Hansen and I.R.McDonald, {\it Theory of simple liquids} 
(Academic Press, London, 1986), 2nd Edition.
\end{thebibliography}
\end{document}